# Review of Apriori Based Algorithms on MapReduce Framework

Sudhakar Singh[1,*], Rakhi Garg[2], P K Mishra[1]

[1] *Department of Computer Science, Faculty of Science, Banaras Hindu University, Varanasi.*
[2] *Mahila Maha Vidyalaya, Banaras Hindu University, Varanasi.*
e-mail: * sudhakarcsbhu@gmail.com

**Abstract**

The Apriori algorithm that mines frequent itemsets is one of the most popular and widely used data mining algorithms. Now days many algorithms have been proposed on parallel and distributed platforms to enhance the performance of Apriori algorithm. They differ from each other on the basis of load balancing technique, memory system, data decomposition technique and data layout used to implement them. The problems with most of the distributed framework are overheads of managing distributed system and lack of high level parallel programming language. Also with grid computing there is always potential chances of node failures which cause multiple re-executions of tasks. These problems can be overcome by the MapReduce framework introduced by Google. MapReduce is an efficient, scalable and simplified programming model for large scale distributed data processing on a large cluster of commodity computers and also used in cloud computing.
In this paper, we present the overview of parallel Apriori algorithm implemented on MapReduce framework. They are categorized on the basis of Map and Reduce functions used to implement them e.g. 1-phase vs. k-phase, I/O of Mapper, Combiner and Reducer, using functionality of Combiner inside Mapper etc. This survey discusses and analyzes the various implementations of Apriori on MapReduce framework on the basis of their distinguishing characteristics. Moreover, it also includes the advantages and limitations of MapReduce framework.

## 1. Introduction

Advances in storage, communication and networking technologies leads to rapid growth of huge volume of data in both scientific and commercial domain. Organizations are intended to gain insightful and precise information from such large volume of data, which can be understandable and process by human brain for decision making. Data mining and knowledge discovery has emerged to extract useful, hidden and unknown patterns and knowledge from large database. Association Rule Mining (ARM) is one of the most important and popular technique of data mining which find interesting correlation or association between set of items or attributes and also frequent patterns in large database [1]. The most typical application of ARM is in market basket analysis which analyzes the purchasing behavior of customers by finding the frequent items purchased together. In addition to the many business application, it is also applicable to bi-informatics, medical diagnosis and text analysis [2].

---

*Corresponding author

Various ARM algorithms have been developed that differs from each other in the way the strategy they used. These strategies are based on candidate generation, without candidate generation, based on equivalence class clustering, maximal hypergraph clique clustering and lattice traversal scheme [3], [4], [5]. When it comes to mine huge volume of data, these algorithms failed to prove scalability and efficiency. The main reasons behind this are the processing capacity, storage capacity and RAM of a single machine [6]. Hence parallel and distributed algorithms are developed to perform large-scale computing in association rule mining on multiple processors. These parallel and distributed algorithms improve the mining performance but also add some overheads like partition of input data, workloads balancing, reduction in communication costs and aggregation of information at local nodes to form the global information. There are various such algorithms developed that addresses these issues in homogeneous computing environment [7], [8], [9], [10]. These traditional parallel and distributed algorithms are not suitable for heterogeneous environment like heterogeneous cluster and grid environment. Consequently, grid based ARM algorithms have been proposed to improve the performance of mining association rule in grid computing environment [11], [12], [13], [14], [15], [16], [17], [18].

The problems with most of the distributed framework are overheads of managing distributed system and lack of high level parallel programming language. Working with large number of computing nodes in cluster or grid, there is always a potential chance of node failures which causes multiple re-executions of tasks. On the other hand, message passing interface (MPI) is the most popular framework for scientific distributed computing but only works with low level language like C and FORTRAN [6], [19], [20]. All these problems can be overcome by the MapReduce framework introduced by Google [21], [22]. MapReduce is a simplified programming model for large scale distributed data processing and also used in cloud computing. Hadoop is an implementation of Google's MapReduce by Apache which is available as open source [23].

In this paper, we have reviewed various Apriori based algorithms on Hadoop MapReduce framework. The rest of the paper is structured as follows. Section 2 briefly describes the association rule mining problem and parallel Apriori algorithm. Section 3 briefly introduces the Hadoop Distributed File System and MapReduce programming model. We analyze and discuss the various MapReduce based Apriori algorithms in Section 4. Section 5 categorizes the advantages and limitations of using MapReduce. Finally, Section 6 concludes the paper and discusses future research.

## 2. Association Rule Mining Algorithm

### 2.1. Problem Statement of ARM

The formal statement of mining association rule as defined in [3] can be stated as- let $I = \{i_1, i_2, ..., i_m\}$ be set of attributes called items, $D = \{T1, T2, ..., Tn\}$ be a set of transactions called database. Each transaction $Ti$ in database $D$ is a set of items called itemset such that $Ti \subseteq I$. Each transaction $Ti$ is uniquely identified by a unique transaction identifier $TID$. Let $X$ be a set of items in $I$, a transaction $Ti$ contains $X$ if $X \subseteq Ti$ and support, $sup(X)$ of an itemset $X$ is the percentage of transactions containing $X$ in the database $D$. An *association rule* is a conditional implication of the form $X \rightarrow Y$ where $X, Y \subset I$ are itemsets and $X \cap Y = \emptyset$. The support, $s$ of the rule $X \rightarrow Y$ is the percentage of transaction in $D$ that contain both $X$ and $Y$, and the confidence, $c$ is the percentage of transaction in $D$ containing $X$ that also contains $Y$ [2], [3], [24]. The problem of mining association rule is to find only interesting rule while pruning all uninteresting rules. Support and confidence are the two interestingness criteria used to measure the strength of association rules.

In short we can say that ARM is a two steps process - (i) Generation of frequent itemsets whose support is greater than or equal to the minimum support threshold set by user from database *D* and (ii) Strong association rules that have confidence greater than or equal to the minimum confidence threshold set by user are generated from these frequent itemsets [2], [3].

*2.7. Parallel Apriori Algorithm*

Apriori is the basic and most popular algorithm proposed by R. Agrawal and R. Srikant [3] for finding frequent itemsets based on candidate generation. Candidates are itemsets containing all frequent itemsets. The name of the algorithm Apriori is based on the Apriori property which states that all nonempty subsets of a frequent itemset must also be frequent [2]. The core step of the algorithm is generation of candidate k-itemsets $C_k$ from frequent (k-1)-itemsets $L_{k-1}$ and it consists of *join* and *prune* actions. In join step, the conditional join of $L_{k-1} \bowtie L_{k-1}$ is assigned to candidate set $C_k$ while prune step reduces the size of $C_k$ using Apriori property [2].

The demerits of the serial algorithms are high I/O cost due to iterative scan of database and large consumption of memory. Many sequential and parallel algorithms have been proposed to improve the performance of Apriori algorithm. Some popular sequential approaches are Hash-based technique, Transaction reduction, Partitioning, Sampling and Dynamic itemset counting (DIC) [25], [26], [27], [28], [29]. R. Agrawal and J. Shafer [7] proposed the three parallel version of Apriori algorithm, Count Distribution (CD), Data Distribution (DD) and Candidate Distribution.

CD and DD algorithms are categorized under data parallelism and task parallelism while candidate distribution algorithm is hybrid of data parallelism and task parallelism [38]. Table 1 briefly describes the CD and DD algorithms along with serial Apriori. Detailed algorithms can be found in [3], [7]. In CD and DD algorithms database is evenly distributed across the processing node such as $D^i$ denotes the local database partition at processor $P^i$ and $C_k^i$ & $L_k^i$ are the candidate k-itemsets and frequent k-itemsets respectively at processor $P^i$.

Table 1: Algorithms: Serial Apriori, CD and DD [3], [7]

| // Serial Apriori | // Count Distribution | // Data Distribution |
|---|---|---|
| 1. Scan the database and generate frequent 1-itemsets $L_1$.<br>2. To generate frequent k-itemset $L_k$ where k ≥ 2<br>Find candidate k-itemsets $C_k$ from $L_{k-1}$ as $C_k$ = conditional joint of $L_{k-1}$ with itself.<br>3. Scan the database again and count the support of candidates in $C_k$.<br>4. Find the frequent k-itemsets $L_k$ from $C_k$ as $L_k$ = all candidates of $C_k$ with minimum support.<br>5. Repeat from step 2 for next iteration. | 1. Candidate 1-itemsets, $C_1$ = I, the set of all items.<br>2. To generate $L_k$, where k ≥ 1 Processor $P^i$ scan the local partition $D^i$ and find the local support counts of candidates in $C_k$<br>3. Processor $P^i$ exchange the local counts with other processors to find global counts of candidates in $C_k$.<br>4. Find the frequent k-itemsets $L_k$ from $C_k$ as $L_k$ = all candidates of $C_k$ with minimum support.<br>5. Generate the candidates $C_{k+1}$ from $L_k$ as in serial Apriori and repeat from step 2 for next iteration. | 1. Candidates are distributed in round robin fashion. Local candidate 1-itemsets, $C_1^i \subseteq I$.<br>2. To generate $L_k$, where k ≥ 1 Processor $P^i$ scans the local $D^i$ to find the local support counts of candidates in $C_k^i$.<br>3. Each processor broadcast their local database and receives from other processors. Global support counts of candidates in $C_k^i$ are calculated by scanning all these local databases.<br>4. Each processor $P^i$ calculates $L_k^i$ from local $C_k^i$ and exchange $L_k^i$ with other processors to get complete $L_k$.<br>5. Generate the candidate $C_{k+1}$ from $L_k$ and partition it across the processors. Repeat from step 2 for next iteration. |

The step 3 is of our interest in the three algorithms described in table 1. CD and DD parallelize the step 3 of serial Apriori using count exchange and dataset exchange respectively. Candidate Distribution algorithm reduces the overhead of synchronization and processor communication in CD and DD [6]. After a certain number of passes determined by heuristic, it divides the candidate itemsets and database across processors to work independently.

## 3. Apache Hadoop MapReduce Framework

Hadoop is a large-scale distributed batch processing infrastructure for parallel processing of big data on large cluster of commodity computers [30]. Hadoop is an open source project of Apache [23] which implemented Google's File System [31] as Hadoop Distributed File System (HDFS) and Google's MapReduce [21] as Hadoop MapReduce programming model.

*3.1. Hadoop Distributed File System*

Hadoop Distributed File System (HDFS) is distributed file system that holds a large volume of data in terabytes or petabytes scale and provides fast and scalable access to such data [30]. It stores files in a replicated manner across different machine to provide fault tolerance and high availability during execution of parallel applications [30].

HDFS is a block-structured file system and breaks a file into fixed size blocks (default block size is 64MB) to store across several machines. Hadoop uses two types of machine working in a master-worker fashion, a *NameNode* as master machine and a number of *DataNodes* as worker machines as shown in Figure 1. The NameNode assigns block *ids* to the blocks of a file and stores metadata (file name, permission, replica, location of each block) of the file system in its main memory providing fast access to this information. DataNodes are the individual machines in the clusters which store and retrieve the replicated blocks of multiple files [30].

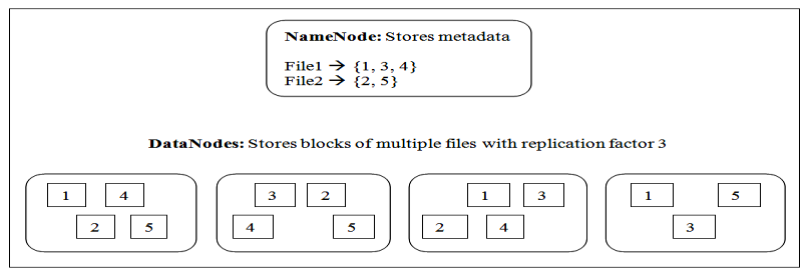

Figure 1: Functionality of NameNode and DataNodes with block replication [30].

*3.2. Hadoop MapReduce*

MapReduce is a parallel programming model designed for parallel processing of large volumes of data by breaking the job into independent tasks across a large number of machines. MapReduce is inspired form the list processing languages e.g. LISP. It uses two list processing idioms: map and reduce. Based on it a MapReduce program consists of two functions *Mapper* and *Reducer* which runs on all machines in a Hadoop cluster. The input and output of these functions must be in form of *(key, value)* pairs [30].

The Mapper takes the input ($k_1$, $v_1$) pairs from HDFS and produces a list of intermediate ($k_2$, $v_2$) pairs. An optional *Combiner* function is applied to reduce communication cost of transferring intermediate outputs of mappers to reducers. Output pairs of mapper are locally sorted and grouped on same key and feed to the combiner to make local sum. The intermediate output pairs of combiners are shuffled and exchanged between machines to group all the pairs with the same key to a single reducer. This is the only one communication step takes place and handle by the Hadoop MapReduce platform. There is no other communication between mappers and reducers take place. The Reducer takes ($k_2$, list ($v_2$)) values as input, make sum of the values in list ($v_2$) and produce new pairs ($k_3$, $v_3$) [30], [32]. Figure 2 illustrates the work flow of MapReduce.

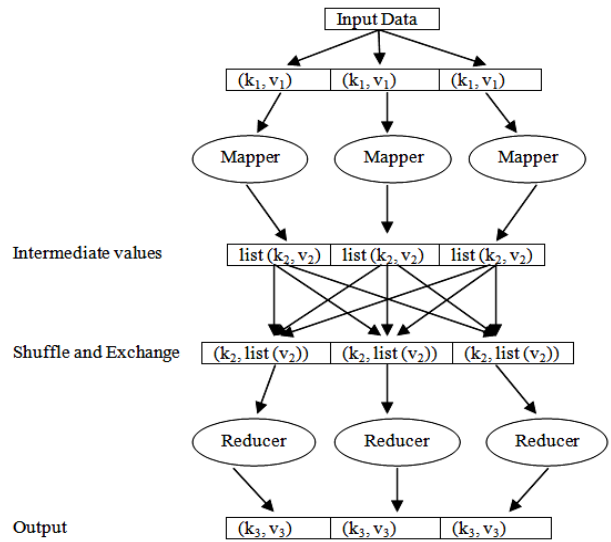

Figure 2: MapReduce Model [30].

MapReduce is a simplified programming model since all the parallelization, inter-machine communication and fault tolerance are handled by run-time system [21].

**4. Apriori Algorithm on Hadoop MapReduce**

To implement an algorithm on MapReduce framework the main tasks are to design two independent map and reduce functions for the algorithm and to convert the datasets in the form of (key, value) pairs. In MapReduce programming, all the mapper and reducer on different machines execute in parallel fashion but the final result is obtained only after the completion of reducer. If algorithm is recursive, then we have to execute multiple phase of map-reduce to get the final result [33].

*4.1. Traditional Apriori to MapReduce Based Apriori*

Apriori algorithm is an iterative process and its two main components are candidate itemsets generation and frequent itemsets generation. In each scan of database, mapper generates local candidates and reducer sums up the local count and results frequent itemsets. The count distribution parallel version of Apriori is best suited on Hadoop whereas to implement data distribution algorithm we have to control the distribution of data which is automatically controlled by Hadoop [6].

The first step of the algorithm is to generate frequent 1-itemsets $L_1$ which is illustrated in Figure 3 by an example. HDFS breaks the transactional database into blocks and distribute to all mappers running on machines. Each transaction is converted to (key, value) pairs where key is the TID and value is the list of items i.e. transaction. Mapper reads one transaction at time and output (key', value') pairs where key' is each item in transaction and value' is 1. The combiner combines the pairs with same key' and makes the local sum of the values for each key'. The output pairs of all combiners are shuffled & exchanged to make the list of values associated with same key', as (key', list (value'')) pairs. Reducers take these pairs and sum up the values of respective keys. Reducers output (key', value''') pairs where key' is item and value''' is the support count ≥ minimum support, of that item [19], [34], [37]. Final frequent 1-itemsets $L_1$ is obtained by merging the output of all reducers.

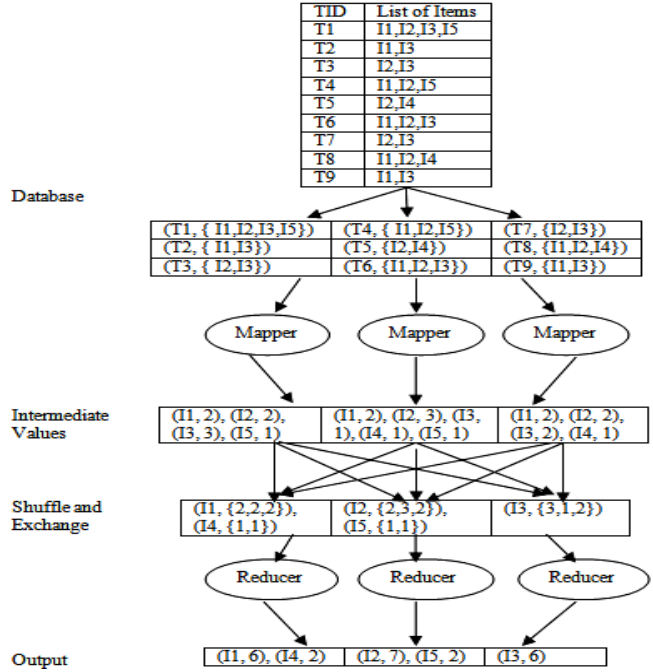

Figure 3: Generation of Frequent 1-itemsets.

To generate frequent k-itemsets $L_k$, each mapper reads frequent itemsets $L_{k-1}$ from previous iteration and generates candidate itemsets $C_k$ from $L_{k-1}$ as in traditional algorithm. A candidate itemset in $C_k$ is selected as key and assigned a value 1, if it is present in the transaction assigned to the mapper. Now we have (key, value) pairs where key is k-itemset and value is 1. All the remaining procedures are the same as generation of $L_1$ [19], [34], [37]. Table 2 depicts the algorithms corresponding to mapper, combiner and reducer for Apriori algorithm.

Table 2: Algorithm: Mapper, Combiner and Reducer [19], [34], [37]

| Mapper (key, value) | Combiner (key, value) | Reducer (key, value) |
|---|---|---|
| // key: TID<br>// value: itemsets in transaction Ti<br>for each transaction Ti assigned to Mapper do<br>  for each itemset in $C_k$ do<br>    if itemset ∈ Ti<br>      output (itemset, 1);<br>    end if<br>  end for<br>end for | // key: itemset<br>// value: list (1)<br>for each itemset do<br>  for each 1 in list (1) of corresponding itemset do<br>    itemset.local_sup + = 1;<br>  end for<br>  output (itemset, itemset.local_sup);<br>end for | // key: itemset<br>// value: list (local_sup)<br>for each itemset do<br>  for each local_sup in list (local_sup) of corresponding itemset do<br>    itemset.sup + = local_sup;<br>  end for<br>  if itemset.sup ≥ minimum support;<br>    output (itemset, itemset.sup);<br>end for |

*4.2. Analysis of Various Proposed Implementations of Apriori on MapReduce*

Various implementations of Apriori on MapReduce framework have been proposed since the inception of MapReduce, introduced by Google. These algorithms can be classified in two categories: 1-phase of map-reduce and k-phase of map-reduce [39]. Also some algorithms used all the three functions mapper, reducer and combiner while some used only mapper and reducer function. Table 3 compares the various proposed algorithms on the basis of some common characteristics.

Table 3: Comparison of Algorithms on Common Characteristics

| Algorithm Proposed by | Map-Reduce Phase | Mapper (key, value) | Combiner (key, value) | Reducer (key, value) |
|---|---|---|---|---|
| X. Y. Yang et al., 2010 [36] | k-phase | Input: key = line no.; value = one row of data<br>Output: key = candidate itemset; value = 1 | Not used | Input: key = candidate itemset; value = list (1)<br>Output: key = itemset; value = support |
| L. Li and M. Zhang, 2011 [37] | 1-phase | Input: key = TID; value = transaction<br>Output: key = itemset; value = 1 | Input: key = itemset; value = list (1)<br>Output: key = itemset; value = local support | Input: key = itemset; value = list (local support)<br>Output: key = itemset; value = support |
| N. Li et al., 2012 [34] | k-phase | Input: key = offset in byte of record; value = string of the content of record<br>Output: key = candidate itemset; value = 1 | Not used | Input: key = candidate itemset; value = list (1)<br>Output: key = candidate itemset; value = support |
| J. Li et al., 2012 [35] | k-phase | Input: key = line no.; value = one line of data<br>Output: key = itemset; value = 1 | Input: key = itemset; value = list (1)<br>Output: key = itemset; value = local support | Input: key = itemset; value = list (local support)<br>Output: key = itemset; value = support |
| M-Y Lin et al., 2012 [19] | k-phase | Input: key = TID; value = itemset in transaction with TID<br>Output: key = itemset; value = 1 | Input: key = itemset; value = list (1)<br>Output: key = itemset; value = local support | Input: key = itemset; value = list (local support)<br>Output: key = itemset; value = support |
| S. Oruganti et al., 2013 [6] | k-phase | Input: key and value are not specified<br>Output: key = candidate itemset; value = 1 | Not used | Input: key = candidate itemset; value = list (1)<br>Output: key = candidate itemset; value = support |
| F. Kovacs and J. Illes, 2013 [33] | k-phase | Input: key = TID; value = itemset in transaction with TID<br>Output: key = itemset; value = local support | Not used | Input: key = itemset; value = list (local support)<br>Output: key = itemset; value = support |

Table 3 summarizes the algorithms with some common characteristics. Now we focus on some of the key features that discriminate these algorithms stated in table 3 as:

*4.2.1. 1-phase vs. k-phase.*

Algorithm based on 1-phase requires only single iteration of map-reduce job to find all frequent itemsets whereas algorithm based on k-phase requires multiple iterations of map-reduce job. Algorithm proposed by L. Li and M. Zhang [37] uses a single iteration of map-reduce job. It generates a set of local candidate itemsets from the input data subset assigned to mapper. Mapper outputs a (itemset, 1) pairs for each candidate itemsets. This is the only algorithm among all the discussed above that uses single map-reduce phase and all the other algorithms are based on k-phase.

*4.2.2. I/O of Mapper, Combiner and Reducer.*

Input (key, value) pairs of mapper of different algorithms are a little different whereas output pairs of all algorithms are similar generated as (itemset, 1) pairs except algorithm [33]. After completion of mapper, itemset is used as a key in both input and output pairs for both combiner and reducer in all algorithms. Combiner function is not used in all proposed algorithms. If combiner is used then list of 1's is passed as a value to combiner and list of local supports is passed as a value to reducer. Some algorithms have not used combiner and passed list of 1's directly to reducer except algorithm [33].

*4.2.3. Using Functionality of Combiner inside Mapper.*

In most of the algorithms mapper outputs each time a (itemset, 1) pair, if itemset is found in the transaction assigned to that mapper. F. Kovacs and J. Illes [33] proposed a different approach. In this algorithm, mapper finally outputs (itemset, itemset.counter) pairs only one time for each itemset where itemset.counter is the local support of itemset. Inside mapper, itemset.counter is incremented each time, if itemset is found in transaction assigned to mapper. In this way, mapper produces output (itemset, local support) which is passed to reducer as (itemset, list (local support)).

*4.2.4. Performance Improvement.*

M-Y Lin et al. [19] proposed three version of Apriori on MapReduce, named Single Pass Counting (SPC), Fixed Passes Combined-counting (FPC) and Dynamic Passes Combined-counting (DPC). SPC is a straight forward implementation of Apriori on MapReduce whereas FPC and DPC implementations improve the performance by reducing the scheduling invocations and waiting time. MapReduce programming model generates scheduling and waiting time overheads since mapper of a map-reduce phase cannot begin until all the reducers of previous phase have finished. Nodes having finished their reducer have to wait for completion of other nodes. FPC combines candidate generation of fixed number of consecutive phases of SPC (generally last phases) into a single map-reduce phase, reducing the number scheduling invocations. DPC dynamically merges the candidates of several consecutive phases to balance the workloads between phases. DPC is better than FPC since DPC dynamically combines several phases while FPC statically combines a fixed number of phases.

*4.2.5. Counting 1 and 2-itemsets in a Single Step.*

F. Kovacs and J. Illes [33] proposed a new method to count the support of 1 and 2-itemsets in one step, without generating candidate 2-itemsets. In this method, a triangular matrix is used to store itemset counters. This matrix contains counts of each 2-itemsets and diagonal of matrix contains counts of 1-itemsets. It is only dependent on number of items and independent of the database size. This method is best suitable for the database having limited number of items. We can directly start the third phase of map-reduce with 2-temset obtained using above method.

*4.2.6. Candidate Generation inside Reducer.*

Mapper outputs the (itemset, 1) pairs and reducer makes the global sum, resulting frequent itemsets. Generation of candidate itemsets from frequent itemsets of previous phase is carried out in mapper. Majority of the algorithms follow this pattern but algorithm proposed by F. Kovacs and J. Illes [33], used reducer instead of mapper, to generate candidate itemsets. Reducer makes the global sum and candidate generation while mapper only generates local support of itemsets.

**5. Advantages and Limitations of MapReduce**

MapReduce is an efficient, flexible and simple model used for large computational problems. Every technique has its advantages and limitations and it depends on for which type of problems we are using it.

*5.1. Advantages*

Some major advantages are as follows which is not only considered for data mining problems but general for all data processing problems.

*5.1.1. Automatic Parallelization, Data Distribution, Workload Balance and Fault Tolerance.*

Underlying run time system of MapReduce parallelizes the execution of mapper and reducer on a number of machines. It partitions the datasets into fixed sized blocks and replicates with some replication factor to provide high availability and zero data loss. It assigns tasks of busy or slower nodes to the idle nodes, balancing the workloads and increasing throughput. MapReduce provides high fault-tolerance by re-executing a crashed task without re-executing the other tasks. It reassigns tasks from failed nodes to idle or active nodes [32]. Programmer is free from such tasks and gives more attention to algorithm.

*5.1.2. Reduced Consumption of Network Bandwidth.*

Hadoop replicates datasets to multiple nodes which allows reading data from local disks, and also writing intermediate data in single copy to local disks, for saving the network bandwidth [21].

*5.1.3. Combination of Computational Power and Distributed Storage.*

Hadoop provides a combined platform for both high computational power and distributed storage system.

*5.1.4. Extremely Scalable.*

MapReduce enables parallel applications to run on a large Hadoop cluster of thousands nodes and process petabyte scale of data [22].

*5.2. Limitations*

Although there are many advantages of MapReduce but it also have some limitations which cannot be ignored. Here we listed some major limitations which are also specific to Apriori algorithm.

*5.2.1. Working on (key, value) Pairs.*

MapReduce model operates only on data structures of type (key, value) pairs. All the input datasets have to be converted into such structure.

*5.2.2. Blocking Operation.*

Result of a map-reduce phase cannot be obtained without completion of reducer. In k-phase of map-reduce, transition to the next phase cannot be possible until all reducers have finished. Consequently, it cannot work on pipeline parallelism [32].

*5.2.3. Implicit Data Distribution.*

Data distribution version of Apriori is not suitable to implement on Hadoop since data distribution is automatically controlled by Hadoop [33]. Count distribution version of Apriori is suitable for Hadoop since it only exchange the count between nodes and does not exchange data [36].

**5. Conclusion**

MapReduce is very lucrative for parallel processing of big data on large cluster of commodity computers. In this paper, we mainly focus on the parallelization of Apriori algorithm on MapReduce framework. The MapReduce computing model is well resembled to the computation of frequent itemsets in Apriori algorithm. We reviewed various proposed approaches to parallelize Apriori on Hadoop distributed framework. They are categorized on the basis of Map and Reduce functions used to implement them e.g. 1-phase vs. k-phase, I/O of Mapper, Combiner and Reducer and using functionality of Combiner inside Mapper etc. Scheduling invocations and waiting time overheads are major bottleneck in performance of algorithms and it is addressed by FPC and DPC techniques. 1 and 2-itemsets are in huge number among all k-candidates so we can handle it separately and input it to the third phase of map-reduce. For this triangular matrix data structure is used to count the support of 1 and 2-itemsets in one step. All these techniques may not be mutually exclusive and some of them can be integrated to increase the performance of resulting algorithm. Although MapReduce is an efficient and scalable platform for big computational problem but it may be difficult to port some problems on such platform.